\documentclass[conference]{IEEEtran}
\IEEEoverridecommandlockouts

\usepackage{algorithmic}
\usepackage{algorithm}
\usepackage{amsmath,amssymb,amsfonts}
\usepackage{graphicx}
\usepackage{textcomp}
\usepackage{xcolor}
\usepackage{multirow}
\usepackage{booktabs}
\usepackage{svg}
\usepackage[numbers, compress]{natbib}
\usepackage{setspace}
\usepackage[utf8]{inputenc}
\usepackage{eso-pic}
\usepackage{hyperref} 
\usepackage{lipsum}   

\def\BibTeX{{\rm B\kern-.05em{\sc i\kern-.025em b}\kern-.08em
    T\kern-.1667em\lower.7ex\hbox{E}\kern-.125emX}}
\begin{document}

\title{Hybrid Disagreement-Diversity Active Learning for Bioacoustic Sound Event Detection \\

}

\author{\IEEEauthorblockN{Shiqi Zhang}
\IEEEauthorblockA{\textit{Audio Research Group} \\
\textit{Tampere University}\\
Tampere, Finland}
\and
\IEEEauthorblockN{Tuomas Virtanen}
\IEEEauthorblockA{\textit{Audio Research Group} \\
\textit{Tampere University}\\
Tampere, Finland}
}
\maketitle

\begin{abstract}

Bioacoustic sound event detection (BioSED) is crucial for biodiversity conservation but faces practical challenges during model development and training: limited amounts of annotated data, sparse events, species diversity, and class imbalance. To address these challenges efficiently with a limited labeling budget, we apply the mismatch-first farthest-traversal (MFFT), an active learning method integrating committee voting disagreement and diversity analysis. We also refine an existing BioSED dataset specifically for evaluating active learning algorithms. Experimental results demonstrate that MFFT achieves a mAP of 68\% when cold-starting and 71\% when warm-starting (which is close to the fully-supervised mAP of 75\%) while using only 2.3\% of the annotations. Notably, MFFT excels in cold-start scenarios and with rare species, which are critical for monitoring endangered species, demonstrating its practical value.

\end{abstract}


\vspace{-2mm}
\section{Introduction}  
\vspace{-2mm}

\textbf{Bioacoustic sound event detection} (\textbf{BioSED})~\cite{1} aims at analyzing the vocal activity of target species in audio recordings. It is  essential for large-scale biodiversity monitoring and ecological conservation efforts. Conventional supervised learning approaches~\cite{1,2,3,4} require extensive annotated data for model training to achieve good performance. This challenge is further compounded when detecting species not present in the training data (unseen species). These limitations motivate the development of labeling-efficient and generalizable methods specifically optimized for bioacoustic monitoring scenarios.

Recent label-efficient learning approaches, including self-supervised~\cite{5, 6}, semi-supervised~\cite{7}, and few-shot~\cite{8, 9} methods, have shown promise in bioacoustic monitoring. However, self-supervised approaches often suffer from a gap between the training process and target tasks. Semi-supervised methods can propagate pseudo-labeling errors, especially in class-imbalanced bioacoustic data. Few-shot learning struggles due to inadequate temporal pattern characterization and cross-device prototype distortion.

These limitations stem from static sampling methodologies on training data that lack adaptive instance prioritization during annotation. This has motivated \textbf{active learning}~\cite{10} — a framework that selects the most informative data for labeling based on uncertainty, diversity, or disagreement, thereby improving model performance while minimizing the number of required annotations.

\AddToShipoutPictureFG*{
  \AtPageLowerLeft{
    \put(\LenToUnit{2cm}, \LenToUnit{2.1cm}){
      \footnotesize
      Repo: \url{https://github.com/TioSisai/bioac-active-learning-eusipco-2025}
    }
  }
}

While active learning illustrates potential in general audio tasks~\cite{11, 12, 13}, its application to BioSED faces unique challenges due to the sparsity and class imbalance of bioacoustic events. Existing BioSED active learning methods often rely on uncertainty, diversity, or hybrid uncertainty-and-diversity-based approaches. Uncertainty methods~\cite{14, 15, 16, 17, 18} seek instances with the lowest model confidence to select the data near the decision boundary, but may lead to volatile behavior and easily get trapped in local minima~\cite{19, 20}. Diversity methods~\cite{21} are effective initially, but may prioritize outliers in later iterations, which might be harmful for the performance~\cite{22}. Even if hybrid uncertainty-and-diversity-based methods~\cite{23, 24, 25, 26, 27, 28} combine the advantages from both, they are still constrained by their reliance on a single model, offering a comparatively restricted information gain in comparison to disagreement-based methods~\cite{29}.

To address these challenges, we adapt the mismatch-first farthest-traversal (MFFT)~\cite{12} framework for BioSED. MFFT leverages prediction disagreement from committee voting to distribute the informativeness dependency of sample selection across multiple models, while employing diversity-aware sampling to alleviate cold-start problems. To our knowledge, this work is the first application of a hybrid disagreement-and-diversity-based active learning approach to BioSED. Experimental results demonstrate that MFFT provides a more efficient and stable solution for active learning-based BioSED in comparison to disagreement or diversity-based baselines.

\noindent In conclusion, our contributions are:

\noindent - We adapt and evaluate MFFT for the BioSED, combining committee disagreement and diversity analysis to improve sample selection over baselines. On the proposed dataset, it achieves a performance close to that of fully annotated supervised learning while using only a small portion of annotations in the cold start scenario.
\newline\noindent - We refine the DCASE 2024 Task 5 dataset~\cite{dcase1, dcase2} for active learning training and evaluation. This multi-label (including common and rare species) benchmark focuses on active learning's efficiency within a limited labeling-budget and extensibility to novel species.
\newline\noindent - We provide a comprehensive analysis of MFFT's performance, including comparisons to baselines (random sampling, pure disagreement/diversity methods) and investigation of its effectiveness in cold-start and rare species detection scenarios.

\section{Methodology}
\label{sec:methodology}

\subsection{Active learning for BioSED}
\label{subsec:albiosed}
\vspace{-1mm}

We process longer audio recordings in fixed-length samples~$x$ for computational efficiency. The goal is to train a model $f_{\theta}$ to give the prediction of the time-series label $y \in \{0, 1\}^{T\times N_{cls}}$ indicating the presence of each of the $N_{cls}$ target classes in $T$ time segments within each sample $x$.

To enhance model training using a minimal labeling budget, \textbf{active learning} is employed. It starts with a dataset of unlabeled audio samples $\mathcal{U}_{0}$, and iteratively selects samples to be labeled. The model will be trained after each iteration using the samples labeled so far. It minimizes annotation effort by selecting the most informative samples, thereby maximizing model performance gain, and formally is defined as:

\boxed{
\begin{aligned}
&\text{Objective:} \quad \min_{\pi} \underbrace{\mathbb{E}_{(x,y)\sim P_{\text{test}}} [\ell(f_\theta(x), y)]}_{\text{Generalization Risk}} + \lambda \cdot \underbrace{\sum_{k=1}^K c(\mathcal{X}_k)}_{\text{Annotation Cost}} \\
&\text{s.t.} \quad \theta = \text{Train}(\mathcal{L}_K), \quad \mathcal{X}_{k+1} = \pi(\mathcal{L}_k, \mathcal{U}_k, f_\theta)
\end{aligned}
}\\
\begin{itemize}
    \item \textbf{Generalization Risk:} The expected loss $\ell$ of the detection model $f_\theta$ is computed over test sample-label pairs $(x, y) \sim P_{\text{test}}$.
    \item \textbf{Annotation Cost:} Cumulative cost $c$ of labeling $K$ sample groups $\mathcal{X}_k$, selected by policy $\pi$. $\lambda > 0$ weights the generalization risk and the annotation cost.
    \item \textbf{Model Training:} Model parameters $\theta$ trained on the set of labeled samples $\mathcal{L}_K=\bigcup_{k=1}^K \mathcal{X}_k$.
    \item \textbf{Sample Selection:} Policy $\pi$ selects the next audio sample group $\mathcal{X}_{k+1}$, informed by $\mathcal{L}_k$ and $f_\theta$, and moves the samples from unlabeled set $\mathcal{U}_k$ to $\mathcal{L}_{k+1}$.
\end{itemize}
In our BioSED framework, we first pre-train an audio encoder and freeze its weights. Raw audio samples are then encoded into temporal vector representations $x_{e}\in\mathbb{R}^{T\times D}$ using this pre-trained encoder, where $D$ is the output dimension of the encoder. The active learning process is applied to train a multi-layer perceptron (MLP)-based classifier head with sigmoid activations to constrain the output values between 0 and 1. It is trained on top of this frozen encoder using the encoded temporal representations $x_{e}$ and corresponding time-series labels $y$ for training, resulting in the trained model $f_{\theta}$. Although the model gets trained and produces output at the resolution of segments, to improve computational efficiency, we utilized temporal max-pooled representations instead of whole segments of a sample in the selecting process.

\subsection{Active Learning Strategies}
\label{sebsec:alstrategies}
\vspace{-1mm}

We evaluate several active learning strategies for BioSED, ranging from a simple random baseline to more sophisticated methods leveraging disagreement and diversity.

\textit{\textbf{1) Random Sampling} (RS)} serves as a baseline, selecting audio samples for annotation uniformly at random from the unlabeled set $\mathcal{U}_k$.  This method requires no model predictions for sample selection. While simple, RS provides a crucial benchmark to assess the effectiveness of more informed active learning strategies in the context of BioSED.

\textit{\textbf{2) Mismatchness Priority} (MP)} leverages the disagreement between a committee of models to identify informative audio samples. In our implementation, the committee consists of two models: the MLP classifier~$f_{\theta}$ in Section~\ref{subsec:albiosed} and a nearest neighbor (NN) classifier~$f_{\theta_{\text{NN}}}$. The MLP classifier is a parametric model trained alongside the active learning process, while NN serves as a non-parametric, simpler model to mitigate potential overfitting problems and enhance committee diversity. In MP experiments, NN classifier naturally produces binary outputs by propagating the label of the nearest sample. To ensure consistency, we threshold the MLP's output using the optimal threshold (in terms of the best mean average precision) derived from the validation set.

For each audio sample $x$ in the unlabeled set $\mathcal{U}_k$, we compute the mismatch score:

\vspace{-2.5mm}
\begin{equation}
m(x) = \big\|\overline{f}_{\theta}(x) - \overline{f}_{\theta_{\text{NN}}}(x)\big\|_{1}
\label{eq:mismatch_rate}
\vspace{-1.5mm}
\end{equation}
 as the count of class-wise discrepancies in the temporal max-pooled predictions from the committee, where $\overline{f}_{\theta}$ and $\overline{f}_{\theta_{\text{NN}}}$ are the temporal max-pooled binary predictions from the MLP and NN models, respectively.




After computing mismatch scores for all samples of the current unlabeled set $\mathcal{U}_k$, the policy $\pi_{\text{MP}}$ selects $|\mathcal{X}_{k+1}|$ (the pre-defined group size) samples with the highest mismatch scores. In cases where many samples exhibit the same mismatch score, samples will be randomly selected from them.

MP prioritizes audio samples where the committee models disagree, hypothesizing these samples are more informative for refining the decision boundaries in BioSED. However, MP can suffer from cold-start issues when models are initially weak.

\textit{\textbf{3) Farthest Traversal} (FT)} focuses on maximizing the diversity of selected audio samples. We compute a cosine distance of the temporal max-pooled representation of each pair of audio samples in the dataset (derived from the frozen pre-trained encoder).

The initial sample is randomly chosen. After that, FT incrementally selects each new sample as:

\vspace{-2.5mm}
\begin{equation}
x = \arg\max_{x \in \mathcal{U}} \min_{s \in \mathcal{S}} d(x, s)
\label{eq:ft}
\vspace{-1.5mm}
\end{equation}
until the number of selected samples reaches the pre-defined group size $|\mathcal{X}_{k+1}|$ of iteration $k$. Here, $\min_{s \in \mathcal{S}} d(x, s)$ is the minimum cosine distance from a sample $x$ to the set $\mathcal{S}$ of already selected samples, and $\mathcal{U}$ means the unlabeled set. After every sample selection, $x$ will be moved from $\mathcal{U}$ to $\mathcal{S}$.

FT promotes diversity in the selected samples, which is beneficial in BioSED to capture the variability in acoustic environments and species vocalizations, especially in the initial stages of active learning. However, in datasets with sparse bioacoustic events and prevalent noise, FT might inadvertently select noisy samples, potentially hindering performance gains.

\textit{\textbf{4) Mismatch-First Farthest-Traversal} (MFFT)} combines the strengths of MP and FT to balance disagreement and diversity throughout the active learning process. Initially, to mitigate the cold-start problem where mismatch scores are unreliable, MFFT prioritizes diversity using the FT strategy. In subsequent iterations, unlabeled audio samples in $\mathcal{U}_k$ are sorted based on their mismatch scores using Eq.~(\ref{eq:mismatch_rate}), and selected according to this ranking. When multiple unlabeled samples have the same mismatch score and their number exceeds the remaining group size, MFFT applies the FT strategy within such a subset. Specifically, it incrementally selects the farthest sample to the selected set $\mathcal{S}$ within this unlabeled subset $\mathcal{U}_{sub}$:

\vspace{-2.5mm}
\begin{equation}
x = \arg\max_{x \in \mathcal{U}_{sub}} \min_{s \in \mathcal{S}} d(x, s)
\vspace{-1.5mm}
\end{equation}
where $\forall x_{i},x_{j}\in \mathcal{U}_{sub},~m(x_{i})=m(x_{j})$, with the remaining notation consistent with Eq.~(\ref{eq:ft}).

MFFT effectively balances the exploitation of informative samples (via mismatch priority) and the exploration of diverse samples (via farthest traversal), offering a robust active learning strategy particularly suited for the challenges of BioSED, including cold-start scenarios and noisy acoustic environments.

\begin{figure*}[t!]
    \vspace{-4mm}
    \centering
    \includegraphics[width=\textwidth]{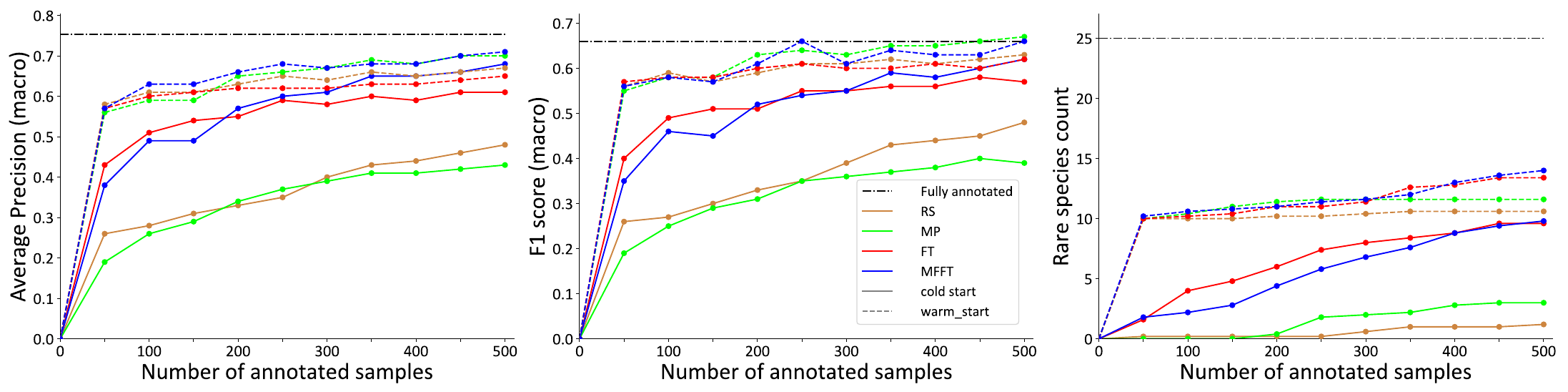} 
    \vspace{-8mm}
    \caption{Performance comparison on test set: mAP (left), F1 score (middle), and rare species count (right) as a function of the number of annotations.}
    \label{fig:example0}
\end{figure*}

\begin{table*}[htbp]  
\centering  
\setlength{\arrayrulewidth}{0.7pt}  
\setlength{\tabcolsep}{6.1pt}  
\begin{tabular}{c|c|cccccccccc|c}  
\hline\hline  
\multicolumn{2}{c|}{} & \multicolumn{11}{c}{Number of annotated samples} \\
\hline  
Methods & Start & 50 & 100 & 150 & 200 & 250 & 300 & 350 & 400 & 450 & 500 & full \\
\hline\hline  
RS   & cold & 26(±4) & 28(±4) & 31(±6) & 33(±5) & 35(±5) & 40(±6) & 43(±7) & 44(±5) & 46(±6) & 48(±6) & 75(±2)\\
\hline  
MP   & cold & 19(±1) & 26(±2) & 29(±1) & 34(±5) & 37(±6) & 39(±7) & 41(±7) & 41(±8) & 42(±8) & 43(±9) & 75(±2)\\
\hline  
FT   & cold & \textbf{43}(±5) & \textbf{51}(±4) & \textbf{54}(±4) & 55(±2) & 59(±2) & 58(±2) & 60(±2) & 59(±2) & 61(±2) & 61(±2) & 75(±2)\\
\hline  
MFFT & cold & 38(±4) & 49(±4) & 49(±5) & \textbf{57}(±4) & \textbf{60}(±6) & \textbf{61}(±6) & \textbf{65}(±5) & \textbf{65}(±4) & \textbf{66}(±6) & \textbf{68}(±3) & 75(±2)\\
\hline\hline  
RS   & warm & \textbf{58}(±4) & 61(±3) & 61(±2) & 63(±2) & 65(±2) & 64(±2) & 66(±3) & 65(±3) & 66(±2) & 67(±1) & 75(±2)\\
\hline  
MP   & warm & 56(±4) & 59(±4) & 59(±2) & 65(±3) & 66(±2) & \textbf{67}(±2) & \textbf{69}(±2) & \textbf{68}(±2) & \textbf{70}(±1) & 70(±3) & 75(±2)\\
\hline  
FT   & warm & 57(±3) & 60(±4) & 61(±4) & 62(±3) & 62(±2) & 62(±1) & 63(±2) & 63(±2) & 64(±1) & 65(±2) & 75(±2)\\
\hline  
\textbf{MFFT} & warm & 57(±3) & \textbf{63}(±5) & \textbf{63}(±3) & \textbf{66}(±3) & \textbf{68}(±2) & \textbf{67}(±1) & 68(±1) & \textbf{68}(±3) & \textbf{70}(±2) & \textbf{71}(±2) & 75(±2)\\
\hline\hline  
\end{tabular}  
\vspace{1mm}
\caption{Average of MAP (in percentages) over five different trials and its standard deviation\\`full' for using all 21,414 labels of samples in the train set}
\vspace{-8mm}
\label{tab:map_results}  
\end{table*}  

\section{Experimental Setup}
\vspace{-1mm}
This section introduces the experimental setup, detailing dataset design, data preprocessing, model selection, and hyperparameter configuration. In this experiment, encoder pretraining follows a standard supervised learning paradigm. For active learning, we use datasets with existing labels but simulate the annotating process by initially ignoring all labels and then iteratively extracting the labels for the selected samples.

\subsection{Dataset Design \& Data Preprocessing}
\label{subsec:dataset_design}
\vspace{-1mm}

To evaluate active learning for BioSED under realistic conditions, we curated a dataset based on the DCASE 2024 Task 5 dataset~\cite{dcase1, dcase2}. This custom dataset was specifically designed to address key challenges in BioSED: extreme class imbalance, rare species, and generalization to novel species not encountered during encoder pre-training.

The original DCASE BioSED development set is pre-partitioned into training and validation sets. The training set includes recordings of 46 species classes, while the validation set contains recordings of 7 other species classes. Both sets are provided with temporal multi-label annotations.

For encoder pre-training, we further divided the original training set into two subsets: 80\% for encoder pre-training and 20\% for validation.  Encoder parameters for active learning were selected based on the best mAP on this encoder validation subset, ensuring the encoder was optimized for robust feature extraction before active learning.

For active learning experiments, the original validation set (unseen species) was partitioned into three subsets: 70\% for active learning training, 15\% for validation, and 15\% for testing. These subsets were used exclusively for evaluating the active learning methodologies detailed in Section \ref{sebsec:alstrategies}. It is important to note that the class distribution within these datasets is markedly unbalanced. For example, the active learning training set contains a substantial proportion of negative samples (only background noise without any target species), accounting for 18680 samples, 83\% of the entire set. In contrast, classes such as `Mosquito', `Red\_Deer', and `Pilot\_whale\_foraging\_buzzes' are represented by 1207 (6\%), 1024 (5\%), and 1089 (5\%) samples, respectively. Furthermore, extremely rare classes, such as `Meerkat\_alarm\_call' and `Meerkat\_move\_call', are represented by only 19 (0.09\%) and 6 (0.03\%) samples, respectively.

Prior to feature extraction, all audio recordings were resampled to 32 kHz. Audio files were split into 10-second samples using a sliding window with a 5-second stride. Each sample was processed using short-time Fourier transform (STFT) with a Hann window (1024 length), 1024 fast Fourier transform length, and 320 hop size. Log mel spectrograms were then computed from the STFT using a mel filterbank with 64 bands, ranging from 50 Hz to 14 kHz, and a minimum amplitude of $1 \times 10^{-10}$. The log mel spectrograms served as input features for pre-training the encoder. Each 10-second sample was then encoded using PANNs~\cite{panns} encoder, which contains 5 pooling layers with a pooling factor of 2. This results in $\frac{2^5\times320}{32000}=0.32$~s time interval between embeddings and $T=32$ embeddings per each sample when the last 8 STFT frames were symmetric-padded during encoding.

During the active learning stage, the pre-trained encoder was frozen and used to extract features, ensuring that the evaluation focused on active learning strategies, independent of feature extraction variability. Segment-level mAP and F1-score were used as primary metrics to quantitatively evaluate active learning performance on this dataset.  This dataset design, incorporating class imbalance, rare species, and unseen species, is crucial for rigorously assessing active learning in realistic BioSED scenarios.

\subsection{Model Selection \& Hyperparameter Configuration}
\vspace{-1mm}

Since the primary focus of this study is to compare the effectiveness of different active learning methods rather than evaluating the feature extraction capabilities of encoders, we adopted the widely-used structure, Pretrained Audio Neural Networks (PANNs)~\cite{panns}, as the encoder. The PANNs encoder was fine-tuned on the pre-training dataset (Section \ref{subsec:dataset_design}) and its parameters were fixed based on validation set performance. For active learning classification, we employed a simple and commonly used Multilayer Perceptron (MLP) classifier.

\subsubsection{\textbf{Encoder Pretraining}}
Encoder pre-training used the following settings: Batch size: 128; Adam optimizer (initial learning rate $5 \times 10^{-4}$, $\text{betas} = [0.9, 0.999]$, $\text{eps} = 1 \times 10^{-8}$, weight decay = 0); ReduceLROnPlateau scheduler (halving LR if validation mAP doesn't improve for 5 epochs, min LR $5 \times 10^{-6}$); Data augmentation (class balancing, frequency band masking up to 2 masks of width 8, Gaussian noise).

\subsubsection{\textbf{Active Learning}}
Active learning was performed iteratively with an annotation budget of 500 samples (2.3\% of the 21,414 sample dataset), simulating a limited labeling-budget. In each iteration, 50 samples (group size) were selected from the unlabeled set using methods from Section~\ref{sebsec:alstrategies}. Samples labeled until each iteration were used to train the model $f_{\theta}$. To isolate active learning effectiveness, all training hyperparameters, except data augmentation, were kept consistent with encoder pre-training. Data augmentation was intentionally excluded in active learning to maintain a direct mapping between encoded features and original audio, preserving sample selection consistency based on unaugmented data.

\subsubsection{\textbf{Fully Annotated Supervised Learning}}
To assess the efficiency of the proposed methodologies by comparing, we trained an MLP classifier, with an architecture identical to that of $f_{\theta}$, via supervised learning utilizing all 21,414 labeled samples from the dataset. Crucially, to ensure a fair comparison, training conditions mirrored those of the active learning experiments: the encoder was kept frozen, and data augmentation was also omitted.

\subsubsection{\textbf{Experimental Reproducibility}}
To minimize the impact of randomness, all experiments were repeated 5 times, and the results show the average value and standard deviation.

\vspace{-2mm}
\subsection{Scenarios}
\vspace{-1mm}

To comprehensively evaluate the performance of active learning methods under various challenging conditions, we designed specific scenarios: cold start, warm start, and rare species detection. Each scenario addresses a unique aspect of real-world bioacoustic data challenges.

\subsubsection{Cold Start \& Warm Start}
`Cold Start' assumes no pre-labeled positive examples, allowing us to evaluate the initialization performance of the active learning methods themselves. In contrast, `Warm Start' provides a small number of labeled positive samples for each class to simulate a more realistic starting point. In this study, we initialized the labeled set with randomly selected $N_{\text{init}} = 5$ positive samples per class, totaling $5\times7=35$ samples. The remaining 15 samples were selected through each active learning strategy. 

\subsubsection{Rare Species Detection}
In the Rare Species Detection scenario, we focus on the model's ability to detect extremely rare species in the dataset. As described in Section \ref{subsec:dataset_design}, the training set for active learning contains only 19 samples of the `Meerkat\_alarm\_call' class and 6 samples of the `Meerkat\_move\_call' class. These two classes are challenging to detect due to their scarcity. For each method, we measure the cumulative number of samples of these rare classes selected (referred to as `rare species count'). A higher count indicates better performance in selecting rare species, reflecting the method's sensitivity to minority classes.

\section{Results \& Discussion}
\vspace{-1mm}

\textbf{Cold-Start and Warm-Start Scenarios:} In both cold-start and warm-start settings, MFFT demonstrates strong performance across annotation budgets (Table \ref{tab:map_results} and Figure \ref{fig:example0}).  In cold-start, FT initially leads at smaller budgets (50-150 samples) due to its diversity-focused sampling, crucial when labeled data is scarce. However, as the number of annotations increases, MFFT surpasses FT, achieving a significantly higher mAP of 68\% at 500 samples—outperforming FT (61\%), RS (48\%) and MP (43\%). In warm-start, MFFT still excels to a top mAP of 71\% at 500 samples, closely followed by MP (70\%) and RS (67\%). Notably, FT’s advantage diminishes in warm-start, suggesting that the benefits of pure diversity decrease when initial labels are available. In such cases, FT even underperforms RS. MP is competitive in warm-start but weak in cold-start with decreased model maturity. These comparisons highlight MFFT's performance, effectively balancing initial diversity exploration (cold-start FT) and later exploitation of informative samples through disagreement (warm-start MP).

\textbf{Method Efficiency, Sensitivity and Trade-offs:} MFFT shows remarkable annotation usage efficiency, achieving 68\% and 71\% in terms of mAP with only 2.3\% (500 samples) of the total dataset in cold and warm-start scenarios, respectively. In contrast, the mAP of the fully annotated supervised learning is 75\%. This highlights MFFT's ability to significantly reduce annotation costs while maintaining high performance. Furthermore, MFFT exhibits superior sensitivity in detecting rare species. As shown in Figure \ref{fig:example0}, MFFT consistently identifies a higher rare species count of samples compared to RS and MP across annotation iterations, indicating its effectiveness in capturing minority classes critical for biodiversity monitoring.  While FT also shows good initial performance in rare species sample detection due to its diversity focus, MFFT's balanced approach leads to better rare species sample discovery and overall detection performance. The results reveal a trade-off between diversity and informativeness in active learning strategies. Diversity-driven methods like FT are beneficial for initial exploration, while informativeness-driven methods like MP become more effective as models mature. MFFT effectively balances these aspects, proving robust across scenarios. However, MFFT's reliance on pairwise distance computations may pose scalability challenges for very large datasets. Future work could explore efficient quantized and soft-valued approximations for enhanced applicability in bioacoustics.

\section{Conclusion}

In this study, we systematically evaluated four active learning strategies, including RS, MP, FT and MFFT, on a bioacoustic dataset derived from DCASE 2024 Task 5. Our experiments focused on challenging scenarios, including extreme class imbalance, rare species detection, and noise exclusion.

The results demonstrate that MFFT's mAP outperforms other methods in both cold-start (68\%) and warm-start (71\%) scenarios, which are close to fully supervised performance (75\%) with only 2.3\% of the total annotations. This highlights its ability to effectively balance informativeness and diversity, making it a robust solution for real-world applications.

\vspace{12pt}

\end{document}